\documentclass[article]{IEEEtran}
\usepackage{graphicx}
\usepackage{caption}
\ifCLASSINFOpdf
\else
\fi
\usepackage[cmex10]{amsmath}
\usepackage{amssymb}
\usepackage{mathrsfs}
\usepackage{verbatim}
\usepackage{graphicx}
\usepackage{epstopdf}
\usepackage{cite}
\DeclareMathOperator{\sinc}{sinc} 
\DeclareMathOperator{\rect}{rect}

\begin{document}
%
\title{Waveform Design using Multi-Tone Feedback Frequency Modulation}
%
%
%

\author{David~A.~Hague,~\IEEEmembership{Member,~IEEE,} and Parker Kuklinski \\ Naval Undersea Warfare Center \\ 1176 Howell St, Newport, RI 02840 \\ Email: david.a.hague@navy.mil}

\maketitle
\begin{abstract}
This paper introduces a waveform design method using Multi-Tone Feedback Frequency Modulation (MT-FFM), a generalization of the single oscillator feedback FM method developed by Tomisawa \cite{tomisawa1981tone}. The MT-FFM utilizes a collection of $K$ harmonically related oscillators each governed by a design parameter $z_k$ which are utilized as a discrete set of parameters that may be modified to generate a richer set of modulation functions than in the single oscillator case.  The resulting modulation function is represented using a form of Kapteyn series composed of Generalized Bessel Functions.  This paper describes the structure of the MT-FFM waveform, derives the Kapteyn series representation of the waveform's modulation function, and demonstrates the design method with a waveform design example.  
\end{abstract}

\IEEEpeerreviewmaketitle

\pagenumbering{gobble}

\section{Introduction}
\label{sec:intro}
%
\IEEEPARstart{C}{ognitive} radar systems have been a topic of great interest in the literature for more than a decade.  These systems seek to utilize information gathered from earlier interactions with the environment to inform the selection of system parameters at the transmit and receive ends that optimize system performance for that given target environment \cite{HaykinI}.  Of the many considerations that go into the design of a cognitive system, the transmit waveform potentially possesses the most design versatility \cite{PrabhuBabu, Aubry}.  Of the many waveform properties that may be adapted, the shape of the waveform's Ambiguity Function (AF) and its zero Doppler counterpart the Auto Correlation Function (ACF) have been of particular interest \cite{WaveformDiversity, Stoica}. 

Recently, the authors have proposed using Multi-Tone Sinusoidal Frequency Modulated (MTSFM) waveforms as a spectrally efficient adaptive waveform design method \cite{HagueI, HagueII}.  The MTSFM's modulation function is represented as a finite Fourier series where the Fourier coefficients are utilized as a discrete set of parameters that may be adjusted to synthesize waveforms with specific user defined properties.  The MTSFM waveform's modulation function is a multi-harmonic generalization of the Sinusoidal FM (SFM), a waveform whose modulation function is composed a single frequency sinusoid which has found use in active sonar \cite{Collins, Ward} as well as FMCW radar \cite{Levanon}.  By utilizing more Fourier harmonics in its modulation function, the MTSFM is able to generate a richer variety of modulation functions which make it well suited to the adaptive waveform design problem.  Another form of FM, known as feedback FM or self modulation \cite{tomisawa1981tone}, involves the output of an oscillator being weighted by a parameter $z$ and fed back into the input of said oscillator.  With the constraint $|z| \leq 1$, the modulation function of this system can be solved in closed form and is represented by a Kapteyn series composed of cylindrical Bessel functions of the first kind.  The Kapteyn series are in turn governed by a lone design parameter $z$.   

The single design parameter $z$ severely restricts the type of modulation functions that can be produced using this method of modulation.  The lack of multiple design parameters in the feedback FM waveform is similar in nature to the SFM with its single harmonic modulation function. This very limitation inspired the development of MTSFM waveforms \cite{HagueI, HagueIII}.  This raises the intriguing question of whether a multi-oscillator feedback system can produce adaptive waveforms in a manner similar to the MTSFM waveform.  This paper introduces a waveform design method using Multi-Tone Feedback Frequency Modulation (MT-FFM), a generalization of the single oscillator feedback FM method developed by \cite{tomisawa1981tone}. The MT-FFM utilizes a collection of $K$ harmonically related oscillators each governed by a design parameter $z_k$ resulting in a discrete set of adjustable parameters.  These additional design parameters allow for generating a richer set of modulation functions than in the single oscillator case.  The resulting modulation function can be represented as a modified Kapteyn series composed of cylindrical Generalized Bessel Functions (GBF) of the first kind \cite{DattoliKapteyn,DattoliBook}.  The GBF based Kapteyn series can be inserted into the MTSFM waveform model to derive closed-form expressions for the MT-FFM waveform's spectrum, AF, and ACF.  This paper describes the structure of the MT-FFM waveform, derives the Kapteyn series representation of the waveform's modulation function, and demonstrates the design method via an illustrative example.  

\section{The Multi-Tone Feedback FM Model}
\label{sec:format}

The FM waveform $s\left(t\right)$ is modeled as a basebanded complex analytic signal with unit energy and duration $T$ defined over the interval $-T/2 \leq t \leq T/2$ expressed as
\begin{equation}
s\left(t\right) = a\left(t\right)e^{j\varphi\left(t\right)}
\label{eq:ComplexExpo}
\end{equation}  
where $a\left(t\right)$ is a real valued and positive amplitude tapering function and $\varphi\left(t\right)$ is the phase modulation function of the waveform.  Unless otherwise specified, the amplitude tapering function $a\left(t\right)$ is a rectangular function normalized by the square root of the waveform's duration $T$ to ensure unit energy expressed as $\rect\left(t/T\right)/\sqrt{T}$.  The rectangularly windowed waveform's instantaneous frequency is solely determined by the modulation function and is expressed as 
\begin{equation}
m\left(t\right) =  \dfrac{1}{2 \pi}\dfrac{\partial \varphi \left( t\right)}{\partial t}.
\label{eq:m}
\end{equation}  
The Ambiguity Function (AF) correlates a waveform's Matched Filter (MF) to its Doppler shifted versions and is defined as \cite{WaveformDiversity}
\begin{equation}
\chi\left(\tau, \nu\right) = \int_{-\infty}^{\infty}s\left(t-\tau/2\right)s^*\left(t+\tau/2\right)e^{j2\pi \nu t} dt
\label{eq:AF}
\end{equation}
where $\nu$ is the doppler frequency shift expressed in Hz.  The waveform's ACF is the zero Doppler cut of the AF expressed as $R\left(\tau\right) = \chi\left(\tau, \nu\right)|_{\nu=0}$.

\subsection{Single-Tone Feedback FM Model}
\label{subsec:ST_Feedback_FM}
This section describes the solution to the single tone feedback FM model.  For simplicity, the modulation function $m\left(t\right)$ is represented as the variable $\phi$.  For the case of a single tone feedback FM system, the output $\phi$ is weighted by a factor $z$ and fed back into the oscillator.  This system is represented by the equation
\begin{equation}
\phi = \sin\left(\omega t + z \phi\right)
\label{eq:st_1}
\end{equation}
where the frequency $\omega$ and modulation index $z$ are both constants with $|z| < 1$ such that \eqref{eq:st_1} defines $\phi$ as a function of $t$.  The solution to this nonlinear equation can be solved in closed form \cite{Benson}.  Kepler's laws of planetary motion state that the angle $\theta$ subtended at the center of the elliptic orbit of a planet measured from the major axis of the ellipse can be represented by the relation 
\begin{equation}
\omega t = \theta - z\sin\theta
\label{eq:st_2}
\end{equation}
where $z$ is the eccentricity of the ellipse which takes on values in the range of $0 \leq z \leq 1$ and $\omega = 2\pi \xi$ is a constant which can be physically interpreted as an angular velocity.  One may arrive at equation \eqref{eq:st_2} via \eqref{eq:st_1} utilizing the substitution $\theta = \omega t + z\phi$.  For convenience, let $\psi = \omega t$.  Since $|z| < 1$, $f\left(\theta\right) = \psi = \theta - z\sin\theta$ is a strictly increasing function of $\theta$.  As $\theta$ is a function of $\psi$, we can write $\psi=f(\theta (\psi))$.  Since $f\left(\theta\right)$ is monotone increasing it is invertible. Additionally, since $f\left(\theta\right)$ is also continuous and bijective, it follows that $\theta (\psi)$ is both continuous and bijective. Using these properties, we have:
\begin{equation}
f(\theta (\psi+2\pi ))=\psi+2\pi =f(\theta (\psi)+2\pi)
\end{equation}
which by bijectivity implies that the function $g(\psi)=\theta (\psi)-\psi$ is periodic in $\psi$ with period $2\pi$. A similar argument shows that $f(\theta (-\psi))=-\psi=f(-\theta (\psi))$, so that $g(\psi)$ is also an odd function of $\psi$. As such, $g(\psi)$ can be written using a Fourier sine series expansion:
\begin{equation}
g(\psi)=z\phi = \sum _{m=1}^\infty b_m\sin{m\psi}.
\label{eq:fourierSeries}
\end{equation}

The resulting Fourier coefficients are calculated as 
\begin{equation}
b_m=\frac{1}{\pi}\int _0^{2\pi} g(\psi)\sin{m\psi}d\psi
\label{eq:fourier1}
\end{equation}
Using integration by parts, \eqref{eq:fourier1} is represented as
\begin{equation}
b_m=\frac{1}{\pi m}\int _0^{2\pi} g'(\psi)\cos{m\psi}d\psi.
\label{eq:fourier2}
\end{equation}
Further, recognizing that $g(\psi)=\theta (\psi)-\psi$, \eqref{eq:fourier2} simplifies to
\begin{equation}
b_m=\frac{1}{\pi m}\int _0^{2\pi} \frac{d\theta}{d\psi}\cos{m\psi}d\psi.
\end{equation}
Utilizing a change of variables and recognizing that $\theta (0)=0$ while $\theta (2\pi )=2\pi$ yields
\begin{equation}
b_m=\frac{1}{\pi m}\int _0^{2\pi} \cos{m\psi}d\theta.
\label{eq:fourier3}
\end{equation}
Substituting $\psi = \theta - \sin\theta $ into \eqref{eq:fourier3} gives
\begin{equation}
b_m=\frac{1}{\pi m}\int _0^{2\pi} \cos\left(m\theta -mz\sin\theta\right)d\theta.
\label{eq:almostBessel}
\end{equation}
The result in \eqref{eq:almostBessel} is the integral representation of the cylindrical Bessel function of the first kind divided by order $m$ which simplifies \eqref{eq:almostBessel} to 
\begin{equation}
b_m = \dfrac{2J_m\{mz\}}{m}.
\label{eq:almostBessel1}
\end{equation}
Substituting \eqref{eq:almostBessel1} into \eqref{eq:fourierSeries} results in the solution
\begin{equation}
\phi = \sum_{m=1}^{\infty}\dfrac{2J_m\{mz\}}{mz}\sin\left(\omega m t\right).
\label{eq:singleToneFeedbackFM}
\end{equation}

\subsection{Multi-Tone Feedback FM Model}
\label{subsec:MT_Feedback_FM}

The MT-FFM system is a system composed of a set of $K$ harmonically related oscillators where the output of the $k^{th}$ oscillator is weighted by a factor $z_k$.  These weighted outputs are then combined to produce an output signal $\phi$ which is then fed back into each oscillator.  The MT-FFM is represented by the equation
\begin{equation}
\phi = \sum_{k=1}^{K}z_k\sin\left(k\left(\omega t + \phi\right)\right).
\label{eq:feedbackFM1}
\end{equation}
To solve for the resulting modulation function $\phi$, we begin by utilizing the substitution $\theta = \omega t + \phi$ in \eqref{eq:feedbackFM1} which results in the expression
\begin{equation}
\omega t=\theta -\sum _{k=1}^K z_k\sin{k\theta}.
\end{equation}
The goal now is to solve for $\theta$ as a function of $t$ following a procedure similar to that of Section \ref{subsec:ST_Feedback_FM}.  Again, for convenience, let $\psi=\omega t$.  By the triangle inequality, we additionally impose the condition that $\sum_k k|z_k| \leq 1$ such that $f(\theta )=\theta -\sum _{k=1}^Kz_k\sin{k\theta}$ is monotone increasing.  Since $f\left(\theta\right)$ is monotone increasing, it is invertible, and since it is also continuous and bijective, it follows that $\theta (\psi)$ is both continuous and bijective.  Using a similar argument to the one in Section \ref{subsec:ST_Feedback_FM}, it also follows that $g(\psi)=\theta (\psi)-\psi$ is a periodic odd-symmetric function and can be represented as a Fourier sine series.

Following the same logic as in Section \ref{subsec:ST_Feedback_FM} leads to an expression similar to \eqref{eq:fourier3}.  Utilizing the substitution $\psi = \theta - \sum_{k=1}^{K}z_k \sin k\theta$ results in the expression
\begin{equation}
b_m=\frac{1}{\pi m}\int _0^{2\pi} \cos\left(m\theta -\sum _{k=1}^K mz_k\sin k\theta\right)d\theta.
\label{eq:almostGBF}
\end{equation}
The integral in \eqref{eq:almostGBF} is the integral expression for the cylindrical GBF of the first kind \cite{DattoliBook} divided by order $m$.  Utilizing this integral identity simplifies \eqref{eq:almostGBF} to
\begin{equation}
b_m=\frac{2}{m}\mathcal{J}_m^{1:K}\{mz_1,...,mz_K\}
\end{equation}
where the GBF takes in the multi-dimensional argument $\{mz_1,...,mz_K\}$.  The GBF is typically written in shorthand notation $\mathcal{J}_m^{1:K}\{mz_k\}$ where $m z_k$ denotes the $k^{th}$ dimensional argument $z_k$ mutliplied by the order $m$.  Finally, the solution is represented as 
\begin{equation}
\phi = \sum _{k=1}^Kz_k\sin{k\theta}=\sum _{m=1}^\infty \frac{2}{m}\mathcal{J}_m^{1:K}\{mz_k\}\sin{\omega m t}.
\label{eq:multiToneFeedbackFM}
\end{equation}
The expression in \eqref{eq:multiToneFeedbackFM} is a form of Kapteyn series for GBFs and is effectively the same as the result obtained in \cite{DattoliKapteyn}.  In this paper we have additionally imposed the condition that $\sum_kk|z_k| \leq 1$ in order for the function $f\left(\theta\right)$ to have a well defined inverse.    

\begin{figure}[ht]
\centering
\includegraphics[width=0.5\textwidth]{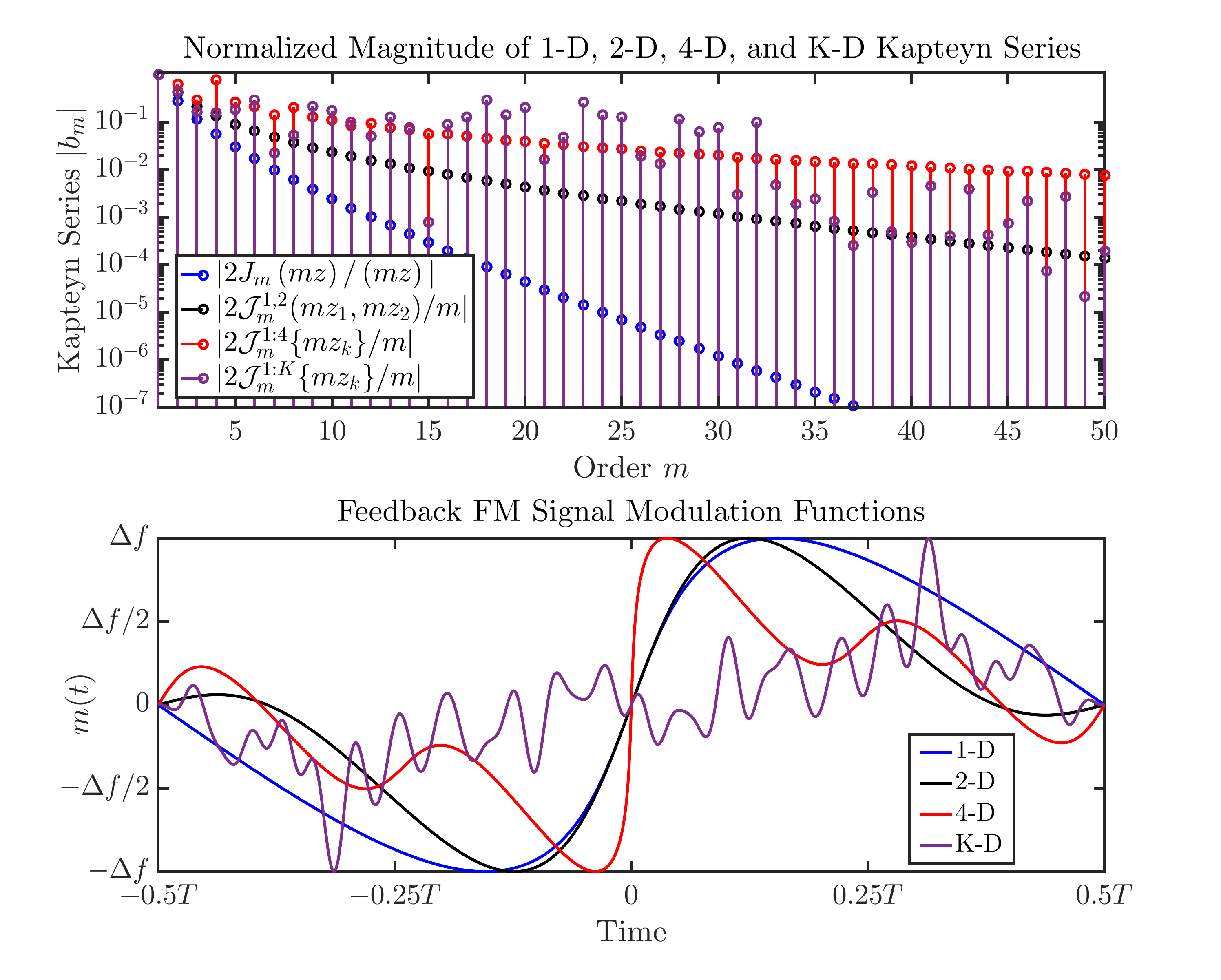}
\caption{Kapteyn series (upper panel) and the resulting modulation functions (bottom panel) of the single oscillator feedback FM and the MT-FFM with $K = 2, 4$, and $32$ design coefficients $z_k$.  Increasing the number of design coefficients allows for creating a richer variety of modulation functions.}
\label{fig:Kapteyn1}
\end{figure}

Figure \ref{fig:Kapteyn1} illustrates the Kapteyn series magnitude and resulting modulation functions for example single oscillator FM and MT-FFMs with a varying number of design coefficients $K$.  Each modulation function is scaled by a factor $A$ such that it is swept through a band of frequencies $\pm \Delta f/2$.  The resulting Kapteyn series, seen in the upper panel, possesses larger values for the lower order harmonics and decay in magnitude at a rate of $1/m$ approaching zero in the limit as $m\rightarrow\infty$.  The single tone feedback FM modulation function shape remains largely unchanged for the permissable values of $z$.  As a result of this, the modulation function's spectral content is dominated by the fundamental harmonic and closely resembles a simple sine wave with some harmonic distortion.  Utilizing the MT-FFM model produces noticeably different results.  As the number of design coefficients $K$ increases, the magnitude of the resulting GBF Kapteyn series coefficients decay less rapidly and even increase briefly before undergoing decay.  The resulting modulation functions from these Kapteyn series now display substantial contributions from higher order harmonics as seen in the bottom panel of Figure \ref{fig:Kapteyn1}.  Additionally, modifying the coefficients $z_k$ can produce Kapteyn series with a wider variety of harmonic content resulting in a richer set of waveform modulation functions compared to the single oscillator feedback FM model.  

\vspace{-1em}
\section{Implementing The MT-FFM Waveform Design Method}
\label{sec:designExample}
This section describes the MT-FFM's time-domain representation, spectrum, and AF.  It also demonstrates the MT-FFM waveform model via a pulse compression waveform design example.  

\subsection{Realizing the MT-FFM Waveform}
\label{subsec:MT_FFM_Model}
Integrating \eqref{eq:multiToneFeedbackFM} and multiplying by $2\pi$ results in the instantaneous phase of the waveform
\begin{equation}
\varphi\left(t\right) = \sum_{m=1}^{\infty}\dfrac{-2AT\mathcal{J}_m^{1:K}\{mz_k\}}{m^2}\cos\left(\frac{2\pi m t}{T}\right)
\label{eq:phiMTFFM}
\end{equation}
where again $A$ is a scaling factor that scales the modulation function to sweep across a band of frequencies $\pm \Delta f/2$.  The transmit waveform $s\left(t\right)$ is realized by inserting \eqref{eq:phiMTFFM} into \eqref{eq:ComplexExpo}.  Utilizing a Jacobi-Anger like expansion for GBFs results in a modified expression for $s\left(t\right)$
\begin{equation}
s\left(t\right) = \dfrac{1}{\sqrt{T}}\sum_{\ell=-\infty}^{\infty}\mathcal{J}_{\ell}^{1:\infty}\{\alpha_m;j^{-m}\} e^{\frac{-j2 \pi \ell t}{T}}
\label{eq:jacobiAnger}
\end{equation}
where $\mathcal{J}_{\ell}^{1:\infty}\{\alpha_m;j^{-m}\}$ is the infinite variable/parameter GBF \cite{DattoliBook} and $\alpha_m = -2AT\mathcal{J}_m^{1:K}\{mz_k\}/m^2$.  We abbreviate the infinite variable/parameter GBF as simply $\mathcal{J}_{\ell}^{1:\infty}$.  The expression for the MT-FFM in \eqref{eq:jacobiAnger} is now in a form similar to that of the MTSFM \cite{HagueII} which allows for deriving closed form expressions for the spectrum and AF.  Following \cite{HagueII}, the spectrum and AF of the MT-FFM are expressed as
\begin{equation}
S\left(f\right) = \sqrt{T}\sum_{\ell=-\infty}^{\infty}\mathcal{J}_{\ell}^{1:\infty}\sinc\left[\pi T\left(f-\frac{\ell}{T}\right)\right],
\end{equation}
\vspace{-1em}
\begin{multline}
\chi\left( \tau, \nu\right) = \left(\frac{T-|\tau|}{T}\right) \sum_{\ell, \ell'}\mathcal{J}_{\ell}^{1:\infty} \left(\mathcal{J}_{\ell'}^{1:\infty}\right)^* e^{-j\frac{\pi\left(\ell+\ell'\right) \tau}{T}} \times \\  \sinc\left[\pi\left(T-|\tau|\right)\left(\nu + \left(\frac{\ell-\ell'}{T}\right)\right)\right].
\label{eq:MT_FFM_NAF}
\end{multline}
The ACF of the MT-FFM is found by setting $\nu = 0$ in \eqref{eq:MT_FFM_NAF}.

\subsection{An Illustrative Example}
\label{subsec:illustrativeExamplel}
The following design example synthesizes a MT-FFM waveform with a Time-Bandwidth Product, expressed as $TBP = T\Delta f$, of $200$ and $K=32$ design coefficients $z_k$.  The design coefficients are initialized using i.i.d Gaussian random variables scaled in magnitude to ensure that $\sum_k k|z_k| \leq 1.0$.  Utilizing coefficients of this type produce a waveform with a thumbtack-like AF \cite{HagueI}.  The goal of this design example is to further refine the waveform to possess reduced ACF sidelobes compared to that of the traditional thumbtack-like waveform.  In order to achieve this, the metric to be optimized is the Integrated Sidelobe Ratio (ISR).  The ISR is defined as the ratio of area $A_{\tau}$ under $|R\left(\tau\right)|^2$ excluding the mainlobe to the area $A_0$ under the mainlobe of $|R\left(\tau\right)|^2$ \cite{WaveformDiversity} and is expressed as 
\begin{equation}
ISR = \left[\dfrac{\int_{\tau_m}^{T}|R\left(\tau\right)|^2 d\tau}{\int_{0}^{\tau_m}|R\left(\tau\right)|^2 d\tau}\right].
\end{equation}
where $\tau_m$ denotes the location in time-delay of the first null of $|R\left(\tau\right)|^2$ and therefore determines the ACF's mainlobe width.

While optimizing the waveform's ISR, we impose two constraints on the design coefficients.  The first constraint is $\sum_k k|z_k| \leq 1.0$ which ensures the waveform design coefficients $z_k$ obey the necessary bounds such that the GBF based Kapteyn series are convergent.  The second constraint is on the RMS bandwidth of the waveform which measures the spread about DC of the waveform's spectrum and is expressed as 
\begin{IEEEeqnarray}{rCl}
\beta_{rms} = 2\pi\left[\int_{-\infty}^{\infty}f^2|S\left(f\right)|^2df\right]^{1/2}.  
\label{eq:RMS1}
\end{IEEEeqnarray}
The RMS bandwidth is inversely proportional to the area under the mainlobe of the waveform's ACF and thus placing constraints on its size consequently places constraints on the size of the mainlobe area/width.  The RMS bandwidth for the MT-FFM waveform, derived in the appendix, is expressed as 
\begin{IEEEeqnarray}{rCl}
\beta_{rms}^2 = 8\pi^2 A \sum_{m=1}^{\infty}\left(\dfrac{\mathcal{J}_m^{1:K}\{mz_k\}}{m}\right)^2 = 2\pi^2 A\sum_{k=1}^{K}z_k^2.  
\label{eq:RMS2}
\end{IEEEeqnarray}

\begin{figure}[ht]
\centering
\includegraphics[width=0.5\textwidth]{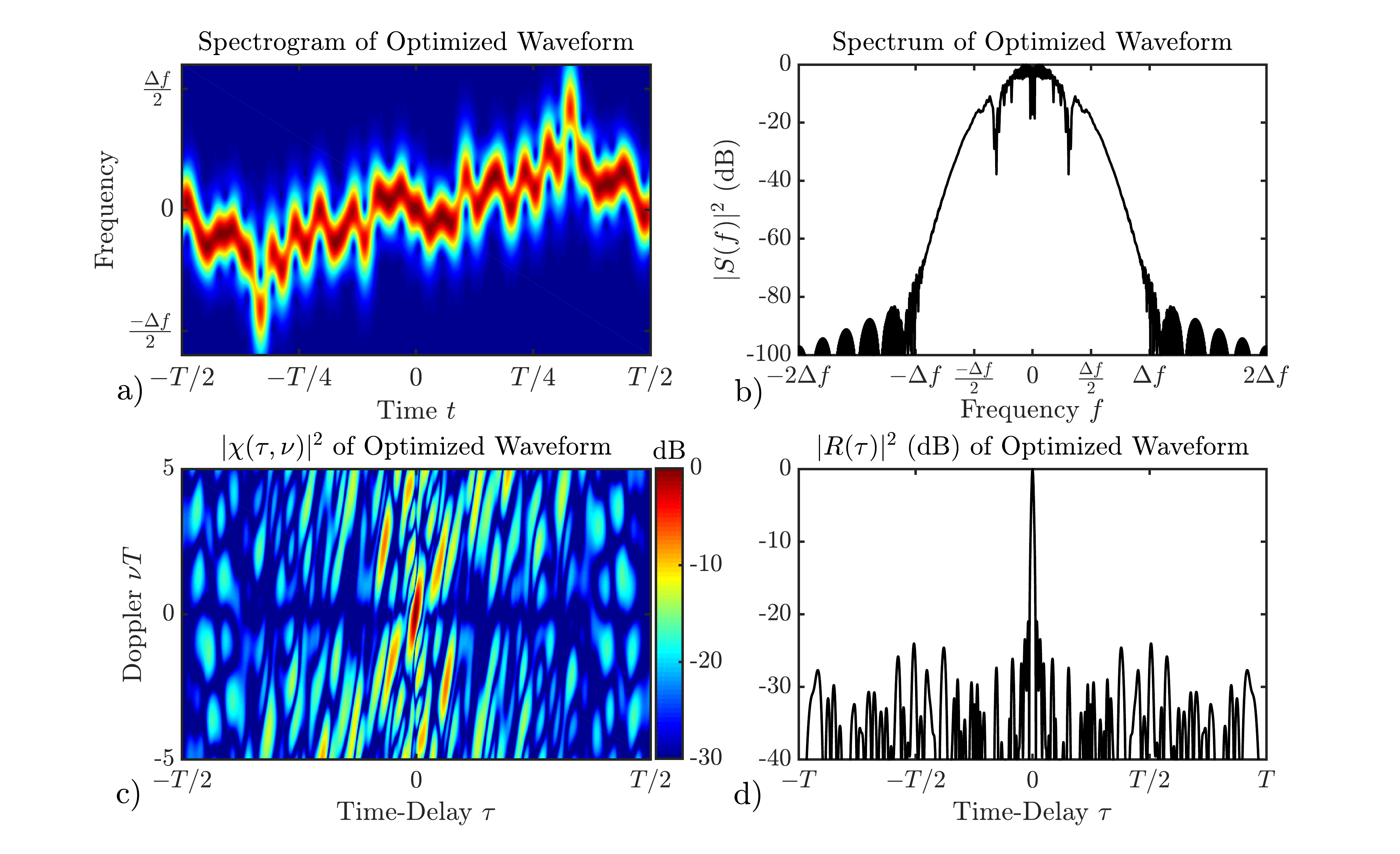}
\caption{Plot of the spectrogram (a), spectrum (b), AF (c), and ACF (d) of the optimized MT-FFM waveform.  The resulting waveform has a mooth modulation function resulting in a spectrally efficient waveform whose AF and ACF shapes may be modified by adapting the $K$ design coefficients $z_k$.}
\label{fig:Kapteyn2}
\end{figure}

The optimization problem is formally expressed as 
\begin{align}
\underset{z_k}{\text{min}}~\left[\dfrac{\int_{\tau_m}^{T}|R\left(\tau\right)|^2 d\tau}{\int_{0}^{\tau_m}|R\left(\tau\right)|^2 d\tau}\right] \text{s.t.}& \sum_k k|z_k| \leq 1\IEEEnonumber\\
& \beta^2_{rms} \geq \left(1-\delta\right)\frac{\pi^2\Delta f}{2}  \IEEEnonumber\\
& \beta^2_{rms} \leq\left(1+\delta\right)\frac{\pi^2\Delta f}{2}
\label{eq:Problem1}
\end{align}
where $\delta = 0.1$ is a unitless parameter and $\pi^2\Delta f/2$ is the initialized waveform's RMS bandwidth.
Figure \ref{fig:Kapteyn2} shows the spectrogram, spectrum, AF, and ACF of the resulting optimized waveform.  The initialized waveform's ISR was $-1.7$ dB and the resulting optimized waveform's ISR was $-9.7$ dB, a $8.0$ dB reduction.  The resulting waveform has a mooth modulation function resulting in a spectrally efficient waveform whose AF is thumbtack-like.

\section{Conclusion}
\label{sec:Conclusion}
The MT-FFM is a generalization of the single oscillator feedback FM method developed by Tomisawa \cite{tomisawa1981tone} which is composed of a collection of $K$ harmonically related oscillators each governed by a design parameter $z_k$.  The resulting waveform's modulation function is represented as a Kapteyn series composed of GBFs.   The design parameters $z_k$ are utilized as a discrete set of adjustable parameters that may be modified to synthesize waveforms with specific AF/ACF characteristics.  In addition to synthesizing adaptive FM waveforms, the MT-FFM model described in this paper may additionally provide insight into the behavior of GBF based Kapteyn series.  Kapteyn series involving the standard 1-D Bessel functions are often encountered in astronomical physics \cite{DattoliKapteyn, Lerche, Tautz} where finding a closed-form expression for the infinite sum of Kapteyn series coefficients in terms of the input argument $z$ can prove insightful \cite{Nikishov}.  It appears that the $n^{th}$ order moments of the MT-FFM waveform's spectrum can be solved in two ways using the left and right sides of \eqref{eq:multiToneFeedbackFM}.  Solving for these moments in this fashion could lead to closed-form expressions for the infinite sum of GBF-based Kapteyn series coefficients in terms of the input arguments $z_k$.  This will be the topic of an upcoming paper.
\appendix[Derivation of the MT-FFM's RMS Bandiwdth]
\label{sec:MTFFM_RMS_Band}
Assuming a rectangularly windowed complex exponential as defined in \eqref{eq:ComplexExpo}, the RMS bandwidth can be defined in the time-domain as
\begin{multline}
\beta_{rms}^2 = \int_{\Omega_t} | \dot{s}\left(t\right)|^2 dt - \left| \int_{\Omega_t} s\left(t \right) \dot{s}^*\left(t\right) dt \right| ^2 \\ = \dfrac{1}{T}\int_{-T/2}^{T/2} \left[\dot{\varphi}\left( t \right)\right]^2 dt - \left| \dfrac{1}{T}\int_{-T/2}^{T/2} j \dot{\varphi}\left(t\right) dt \right| ^2
\label{eq:brms_1}
\end{multline}
where $\dot{\varphi}\left( t \right)$ is the first time derivative of the instantaneous phase equal to $2\pi m\left(t\right)$.  The second term in \eqref{eq:brms_1} is the first spectral moment squared $\langle f\rangle^2$ and is zero since $\dot{\varphi}\left(t\right)$ for the MT-FFM is an odd-symmetric function.  Thus, we are left with only the first integral term to solve for.  There are two ways to solve for the MT-FFM's RMS bandwidth.  The first involves inserting $\dot{\varphi}\left( t \right)$ from \eqref{eq:phiMTFFM} into \eqref{eq:brms_1} and solving the integral.  The second involves inserting the equivalent expression for $\dot{\varphi}\left(t\right)$ as a function of $\theta$ as seen in \eqref{eq:multiToneFeedbackFM} and solving the integral.  

Inserting \eqref{eq:phiMTFFM} into \eqref{eq:brms_1} yields the expression 
\begin{multline}
\frac{4\pi^2A^2}{T}\int_{-T/2}^{T/2}\left[\sum_{m=1}^{\infty}\dfrac{2 \mathcal{J}_m\{m z_k\}}{m}\sin\left(\frac{2\pi m t}{T}\right)  \right]^2 dt \\ = \frac{4\pi^2A^2}{T}\int_{-T/2}^{T/2}\sum_{m,n}\left(\dfrac{2 \mathcal{J}_m\{m z_k\}}{m}\right)\left(\dfrac{2 \mathcal{J}_n\{n z_k\}}{n}\right) \times \\ \sin\left(\frac{2\pi m t}{T}\right)\sin\left(\frac{2\pi n t}{T}\right)   dt .
\label{eq:brms_2}
\end{multline}
Due to the orthogonality of the sine harmonics, the only non-zero terms in \eqref{eq:brms_2} occur when $m=n$.  This simplifies the expression to 
\begin{multline}
\left(\dfrac{4\pi^2A^2}{T}\right)\sum_{m=1}^{\infty}\left(\dfrac{2 \mathcal{J}_m\{m z_k\}}{m}\right)^2\int_{-T/2}^{T/2}\sin^2\left(\frac{2\pi m t}{T}\right) dt.
\end{multline} 
Utilizing the trigonometric identity $\sin^2\theta = \frac{1-\cos2\theta}{2}$, the integral evaluates to $T/2$.  The resulting first expression for $\beta_{rms}^2$ is then expressed as 
\begin{IEEEeqnarray}{rCl}
\beta_{rms}^2 = 8\pi^2 A \sum_{m=1}^{\infty}\left(\dfrac{\mathcal{J}_m^{1:K}\{mz_k\}}{m}\right)^2.
\label{eq:brms_final1}
\end{IEEEeqnarray}

Inserting the equivalent expression from \eqref{eq:multiToneFeedbackFM} into \eqref{eq:brms_1} results in the expression 
\begin{IEEEeqnarray}{rCl}
\frac{1}{2\pi}\int_{-\pi}^{\pi}\left[2\pi A\sum_{k=1}^{K}z_k \sin\theta\right]^2d\theta.
\end{IEEEeqnarray}
Again expanding the square and utilizing the orthogonality of the sine harmonics and utilizing the trigonometric identity $\sin^2\theta = \frac{1-\cos2\theta}{2}$ results in the expression 
\begin{IEEEeqnarray}{rCl}
2\pi A^2 \sum_{k=1}^{K}z_k^2\int_{-\pi}^{\pi}\left(\dfrac{1-\cos2\theta}{2}\right)d\theta.
\end{IEEEeqnarray}
Finally, solving the integral results in the second expression for $\beta_{rms}^2$
\begin{IEEEeqnarray}{rCl}
\beta_{rms}^2 = 2\pi^2 A^2 \sum_{k=1}^{K}z_k^2.
\label{eq:brms_final2}
\end{IEEEeqnarray}
Additionally, the results in \eqref{eq:brms_final1} and \eqref{eq:brms_final2} can be set equal to each other yield an identity for the GBF based Kapteyn series
\begin{IEEEeqnarray}{rCl}
\sum_{m=1}^{\infty}\left(\dfrac{2\mathcal{J}_m^{1:K}\{mz_k\}}{m}\right)^2 = \sum_{k=1}^{K}z_k^2.
\label{eq:Kapteyn_Identity}
\end{IEEEeqnarray}
Reassuringly, for the special case of a 1-D Bessel function (i.e, when $K = 1$), the expression in \eqref{eq:Kapteyn_Identity} collapses back to the known Nielsen formula \cite{Nikishov}
\begin{IEEEeqnarray}{rCl}
\sum_{m=1}^{\infty}\left(\dfrac{J_m\left(mz\right)}{m}\right)^2 = \frac{z^2}{4}.
\label{eq:Nielsen}
\end{IEEEeqnarray}
\section*{Acknowledgment}
This research was funded by the Office of Naval Research (ONR) grant N0001418WX00696 and the internal investment program at the Naval Undersea Warfare Center Division Newport.
\newpage


\end{document}